\documentclass[conference]{IEEEtran}
\usepackage{ifpdf}

\usepackage{cite}

\ifCLASSINFOpdf
   \usepackage[pdftex]{graphicx}
   \graphicspath{{./figs/pdf/}{./figs/jpeg/}}
   \DeclareGraphicsExtensions{.pdf,.jpeg,.png}
\else
   \usepackage[dvips]{graphicx}
   \graphicspath{{./figs/eps/}}
   \DeclareGraphicsExtensions{.eps}
\fi
\usepackage[cmex10]{amsmath,mathtools}
\usepackage{array}

\ifCLASSOPTIONcompsoc
  \usepackage[caption=false,font=normalsize,labelfont=sf,textfont=sf]{subfig}
\else
  \usepackage[caption=false,font=footnotesize]{subfig}
\fi
\usepackage{fixltx2e}
\usepackage{longtable}          
\usepackage{multirow}
\usepackage{framed}
\usepackage[table]{xcolor}
\usepackage{amssymb}
\usepackage{enumerate}
\definecolor{CorCol}{rgb}{0,0,0}

\begin{document}
\allowdisplaybreaks
%
\title{A New Reduction Scheme for Gaussian Sum Filters}
%
%
%


\author{\IEEEauthorblockN{Leila~Pishdad}
\IEEEauthorblockA{Electrical and Computer Engineering Department\\
McGill University\\
Montreal, Canada\\
Email: leila.pishdad@mail.mcgill.ca}
\and
\IEEEauthorblockN{Fabrice~Labeau}
\IEEEauthorblockA{Electrical and Computer Engineering Department\\
McGill University\\
Montreal, Canada\\
Email: fabrice.labeau@mcgill.ca}}

\maketitle

\begin{abstract}
In many signal processing applications it is required to estimate the unobservable state of a dynamic system from its noisy measurements. For linear dynamic systems with Gaussian Mixture (GM) noise distributions, Gaussian Sum Filters (GSF) provide the MMSE state estimate by tracking the GM posterior. However, since the number of the clusters of the GM posterior grows exponentially over time, suitable reduction schemes need to be used to maintain the size of the bank in GSF. In this work we propose a low computational complexity reduction scheme which uses an initial state estimation to find the active noise clusters and removes all the others. Since the performance of our proposed method relies on the accuracy of the initial state estimation, we also propose five methods for finding this estimation. We provide simulation results showing that with suitable choice of the initial state estimation (based on the shape of the noise models), our proposed reduction scheme provides better state estimations both in terms of accuracy and precision when compared with other reduction methods. 
\end{abstract}
%
%
%
%
%
%
%
%
%
%
\section{Introduction}
In many signal processing applications, we require to estimate the inherent state of the system from its observable noisy measurements. Bayesian tracking techniques can be used for this purpose, by providing an approximation of the \textit{posterior} distribution or the \textit{belief function}, the conditional probability density function of the state given the measurements. For linear dynamic state space systems with Gaussian noise, Kalman filter can optimally estimate the Gaussian posterior~\cite{ho_bayesian_1964,anderson1979optimal} by tracking its sufficient statistics, i.e. the mean and the covariance matrix. Additionally, since the Minimum Mean-Square Error (MMSE) estimator is the expected value of the posterior~\cite{poor_introduction_1994,bar2001estimation}, the approximated mean is the MMSE state estimator~\cite{anderson1979optimal}. However, in many applications the noise processes are multimodal and cannot be represented by a Gaussian distribution. 

Gaussian Mixtures (GM) can be used to approximate any distribution as closely as desired\footnote{The parameters can be chosen such that the integral of the approximation error over the sample space is as small as desired.}~\cite[Chapter8; Lemma 4.1]{anderson1979optimal}, and they provide an asymptotically unbiased approximation~\cite{1962KernelDensityEstimation}. Moreover, since a GM distribution is conditionally Gaussian, by modeling the non-Gaussian distributions as GMs, a closed-form expression for the posterior can be evaluated analytically. Specifically, using the \textit{multiple model}  approach~\cite{bar2001estimation}, the GM posterior is viewed as a conditionally Gaussian distribution where each mixand can be optimally tracked by a Kalman filter. Hence, the GM posterior can be approximated by a bank of Kalman filters or the Gaussian Sum Filter (GSF). The mean of this pdf, is the expected value of the posterior, hence the MMSE state estimator~\cite{ackerson_state_1970,tugnait_adaptive_1980,bilik_mmse-based_2010}. 
However, with GM noise distributions, the number of mixands in the posterior grows exponentially and reduction schemes must be used to maintain the size of the bank in GSF. One of the most commonly used methods for reduction is simply removing the clusters with smaller weights and only keeping the mixands with larger weights. This is the approach used in~\cite{alspach_nonlinear_1972,sorenson_recursive_1971,pishdad_approximate_2014} and it is attractive due to its low computational complexity. Another reduction scheme based on removing smaller mixands, can be similar to the resampling algorithm in particle filters. This is the approach used in~\cite{GSPFkotecha_gaussian_2003,andrieu_particle_2002}. Alternatively, in~\cite{ali-loytty_box_2010} a \textit{forgetting and merging} algorithm is proposed, where the mixands with weights smaller than a threshold are removed, and the clusters  with close enough moments are merged. In the most extreme case of removing clusters, the GM distribution can be viewed as a single Gaussian conditioned on a latent variable which determines the mode or the cluster. Hence, by tracking the latent variable, only the corresponding cluster of the GM is kept~\cite{chen_mixture_2000,sun_mixture_2004} and the other mixands are simply removed.

In~\cite{bar2001estimation,GPFkotecha_gaussian_2003,morelande_manoeuvring_2005,djuric_density_2004,bolic_study_2010} instead of removing any clusters, they are all merged to a single Gaussian distribution at the end of each iteration. For higher order models, where the mode-conditioning also includes the history of the state\footnote{For instance in Generalized Pseudo Bayesian estimator of second order (GPB2) the mode conditioning is on the current and previous state.}, Interacting Multiple Models (IMM) can be used to merge the clusters by starting from different initial conditions for each filter in the bank. IMM is less computationally complex when compared \textcolor{CorCol}{with Generalized Pseudo Bayesian estimator of second order (GPB2)}, because unlike GPB2 which requires \(r^2\) filters for \(r\) initial conditions, IMM requires only \(r\) filters at each iteration. Hence, it has been widely used for second order models~\cite{bar2001estimation,morelande_manoeuvring_2005,daeipour_interacting_1995}. Alternatively, the merging of the clusters can be done by taking the shape of the approximated distribution into account. For instance, in~\cite{faubel_split_2009} the clusters are merged in the unlikely regions of the distribution, and they are split in the likely regions. Hence, the mixture reflects the distribution more accurately using less number of mixands.

Another class of reduction algorithms provide solutions by minimizing a cost function, which can be based on Kullback-Leibler divergence~\cite{schoenberg_posterior_2012,bilik_mmse-based_2010}, a least squares error function~\cite{kemouche_gmm_2010}, or an integral squares error~\cite{maybeck_multiple_2005}. Other forms of cost functions can also be used for this purpose. However, these schemes usually suffer from high computational complexity as they use numerical methods for minimizing the cost function.

Alternatively, Expectation Maximization (EM) algorithm can be used to simultaneously predict and reduce the Gaussian Mixture~\cite{huber_efficient_2007,bilik_maneuvering_2010,wang_novel_2012,bilik_optimal_2005}, e.g. by running the EM algorithm on synthetically generated data~\cite{bilik_optimal_2005,bilik_maneuvering_2010}.

Among the reduction schemes mentioned above, the methods that rely on removing clusters are the least computationally complex. This is due to two reasons: 1) Removing the clusters is less computationally complex than minimizing a cost function, EM algorithms or merging all the clusters; 2) If the removing procedure is applied before evaluating the parameters of all filters, computational resources can be saved. Moreover, it is worth noting that at each iteration only one model is \text{active}, and if that model is detected correctly, the lower bound for MMSE can be reached~\cite{flam_mmse_2012}. Hence, in this work we propose a reduction scheme based on removing clusters, which aims at finding the active model or cluster. The proposed method uses an initial state estimation to find the cluster in the posterior which maximizes the a posteriori probability of the noise parameters. \textcolor{CorCol}{This work is different from our previous work~\cite{pishdad_approximate_2014}, in which we modify the cluster means such that the estimator is more robust to removing clusters when compared with GSF. Specifically, in this work, we rely on the cluster means evaluated by GSF to find the active model rather than modifying the cluster means as in~\cite{pishdad_approximate_2014}.}

The major difference between our method and the other reduction schemes is its computational complexity. Specifically, in our proposed method, instead of using computationally complex algorithms to improve the estimation accuracy and precision, we rely on simple comparisons to determine the active model. However, since this method depends on the initial estimation to determine the noise clusters, the accuracy of this estimation can affect the performance of the proposed reduction scheme. Hence, we propose and use different low computational complexity initial estimates and compare the performance of the proposed scheme for each case.

The rest of this paper is organized as follows: In Section~\ref{sec:Notations} the system model is defined and the notations used throughout the paper are introduced. Next, in Section~\ref{sec:GSF}, we provide the details of GSF. In Section~\ref{sec:myCont}, we present our proposed reduction scheme. The simulation results are provided in Section~\ref{sec:SimResults}. Finally, in Section~\ref{sec:Conclusion} we provide concluding remarks. 

\section{System Model}
\label{sec:Notations}
We consider a discrete-time linear dynamic state space system with the following process and measurement equations:
\begin{align}
\label{eq:process}
\mathbf{x}_{k} & = F_{k}\mathbf{x}_{k-1}+\mathbf{v}_{k},\\
\label{eq:measurement}
\mathbf{z}_k & = H_k\mathbf{x}_{k} + \mathbf{w}_k,
\end{align}
where \(\{\mathbf{x}_{k},k \in  \mathbb{N}\}\) and \(\{\mathbf{z}_{k},k \in  \mathbb{N}\}\) are the state and measurement sequences, of dimensions \(n_x\) and \(n_z\), respectively. The matrix \(F_{k}\) which describes the linear relationship between the previous and current state is known and is of size \(n_x \times n_x\). The linear relationship between the current measurement and the current state is described with the matrix \(H_{k}\) of size \(n_z \times n_x\). The process noise \(\{\mathbf{v}_{k},k \in  \mathbb{N}\}\), also of dimension \(n_x\), is an i.i.d. random vector sequence from a GM distribution with \(C_{v_k}\) clusters, \(\left\lbrace \mathbf{u}^i_k,1 \leq i \leq C_{v_k}\right\rbrace  \) cluster means, \(\left\lbrace Q^i_k,1 \leq i \leq C_{v_k}\right\rbrace  \) cluster covariance matrices and \(\left\lbrace w^i_k,1 \leq i \leq C_{v_k}\right\rbrace\) the mixing coefficients. Hence,
\begin{align}
\label{eq:priorDist}
p \left(\mathbf{v}_k\right) \approx \sum \limits _{i=1}^{C_{v_k}} w^i_k \mathcal{N}\left( \mathbf{v}_k; \mathbf{u}^i_k, Q^i_k \right), 
\end{align}
where \(\sum \limits _{i=1}^{C_{v_k}} w^i_k=1\), and \(\mathcal{N}\left( \mathbf{x};\boldsymbol{\mu},\Sigma \right) \) represents a Gaussian distribution with argument \(\mathbf{x}\), mean \(\boldsymbol{\mu}\), and covariance matrix \(\Sigma\). The measurement noise \(\{\mathbf{w}_{k},k \in  \mathbb{N}\}\) is an i.i.d. random vector sequence with the pdf  \(p \left( \mathbf{w}_{k} \right)\), and it is independent from the process noise. Assuming a GM measurement noise we have:
\begin{align}
\label{eq:likelihoodDist}
p \left(\mathbf{w}_k\right) \approx \sum \limits _{j=1}^{C_{w_k}} p^j_k \mathcal{N}\left( \mathbf{w}_k; {\mathbf{b}}^j_k, R^j_k \right),
\end{align}
where, \(C_{w_k}\) is the number of clusters of the GM model with coefficients, \(\left\lbrace p^j_k,1 \leq j \leq C_{w_k}\right\rbrace  \) and \(\sum \limits _{i=1}^{C_{w_k}} p^j_k=1\). The mean and covariance matrix of cluster \( j,1 \leq j \leq C_{w_k}\) are \({\mathbf{b}}^j_k\) and \(R^j_k\), respectively.

\section{Gaussian Sum Filters}
\label{sec:GSF}
In this section we provide an overview of GSF for a linear dynamic system with GM process and measurement noise. In the following, we assume that at each iteration the previous posterior, \(p \left(\mathbf{x}_{k-1} | \mathbf{z}_{1:k-1}\right)\) is Gaussian. This is equivalent to considering a first-order system where at the end of each iteration the number of clusters in the posterior is reduced to one.

Using the GM noise distributions in \eqref{eq:priorDist}--\eqref{eq:likelihoodDist}, the posterior can be partitioned as follows:
\begin{align}
\label{eq:GSFmergedposterior}
p \left(\mathbf{x}_k | \mathbf{z}_{1:k}\right) & = \sum \limits _{i,j} p \left(\mathbf{x}_k | \mathbf{z}_{1:k},M_{k}^{ij}\right)p \left(M_{k}^{ij}|\mathbf{z}_{1:k}\right),
\end{align}
where \(\left\lbrace M_{k}^{ij}; 1 \leq i \leq {C_{v_k}}, 1\leq j \leq {C_{w_k}} \right\rbrace\) represents the \textit{Model} corresponding to cluster \(i\) in the process noise distribution, and cluster \(j\) in the measurement noise distribution. Hence, conditioning on \(M^{ij}\), the posterior is Gaussian and its sufficient statistics can optimally be tracked with a \textit{mode-matched} Kalman filter. This can be written as:
\begin{align}
\label{eq:individualKFposterior}
p \left( \mathbf{x}_k|M_{k}^{ij},\mathbf{z}_{1:k} \right) & = \mathcal{N} \left(\mathbf{x}_k; \mathsf{\hat{\boldsymbol{x}}}_{k|k}^{ij}, \mathsf{P}_{k|k}^{ij}\right),
\end{align}
where \(\mathsf{\hat{\boldsymbol{x}}}_{k|k}^{ij}\) and \(\mathsf{P}_{k|k}^{ij}\) are the mean and covariance matrix of the model-conditioned posterior, approximated by the mode-matched Kalman filter~\cite{ho_bayesian_1964,bar2001estimation}. Hence, defining
\begin{align}
\mu_k^{ij} & \triangleq  p \left(M_{k}^{ij}|\mathbf{z}_{1:k}\right),
\end{align}
we can write the posterior in \eqref{eq:GSFmergedposterior} as a Gaussian Mixture, with \({C_{v_k}}\times{C_{w_k}}\) clusters:
\begin{align}
\label{eq:GSFmergedposteriorSimple}
p \left(\mathbf{x}_k | \mathbf{z}_{1:k}\right) & = \sum \limits _{i,j} \mu_k^{ij} \mathcal{N} \left(\mathbf{x}_k; \mathsf{\hat{\boldsymbol{x}}}_{k|k}^{ij}, \mathsf{P}_{k|k}^{ij}\right).
\end{align}
Using the assumption that the current model is independent from the previous model\footnote{This assumption can be easily relaxed.}, 
we have~\cite{ho_bayesian_1964,bar2001estimation}:
\begin{align}
\label{eq:muijComplete}
\mu_k^{ij} = \frac{w^i_kp^j_k\mathcal{N}\left(\mathbf{z}_k; \mathbf{\hat{z}}_{k}^{ij}, S_k^{ij}\right)}{\sum \limits _{lm} w^l_kp^m_k\mathcal{N}\left(\mathbf{z}_k; \mathbf{\hat{z}}_{k}^{lm}, S_k^{lm}\right)}.
\end{align}
Based on \eqref{eq:GSFmergedposteriorSimple}, we can see that if no reduction scheme is used, the number of the mixands of the GM posterior will increase exponentially over time~\cite{ackerson_state_1970}. This is due to the fact that for each cluster in the previous posterior, we need to consider \({C_{v_k}}\times{C_{w_k}}\) models. 

\subsection{Reduction Schemes}
In this section we provide the details of two most commonly used reduction schemes: 1) Merging all the clusters to one; 2) Removing all the clusters but the one with the most significant weight. These two methods are more commonly used than the others  due to their lower computational complexity.  
 
The first method, i.e. merging all the clusters to one, which is used in~\cite{bar2001estimation,GPFkotecha_gaussian_2003,morelande_manoeuvring_2005,djuric_density_2004,bolic_study_2010}, is equivalent to fitting a single Gaussian distribution to the GM posterior. Hence, the moment-matched Gaussian distribution will have the following mean and covariance matrix:
\begin{align}
\label{eq:mergedState}
\mathsf{\hat{\boldsymbol{x}}}_{k|k} = &\sum \limits _{ij} \mu_{k}^{ij}\mathsf{\hat{\boldsymbol{x}}}_{k|k}^{ij},
\\ \label{eq:mergedCov}\mathsf{P}_{k|k}  
= & \sum \limits _{ij}\mu_{k}^{ij} \left( \mathsf{P}_{k|k}^{ij} + \mathsf{\hat{\boldsymbol{x}}}_{k|k}^{ij} \mathsf{\hat{\boldsymbol{x}}}{_{k|k}^{ij}}^T \right)- \mathsf{\hat{\boldsymbol{x}}}_{k|k}\mathsf{\hat{\boldsymbol{x}}}_{k|k}^T.
\end{align} 

Alternatively, a hard decision can be made about the active cluster, i.e. one active cluster is determined and the others are simply removed. This is the approach used in~\cite{alspach_nonlinear_1972,sorenson_recursive_1971,chen_mixture_2000,sun_mixture_2004}. The active model can be simply chosen by using the weights of the clusters. Using this method, if 
\begin{align}
\label{eq:determineActive}
\mu_k^{ij} = \max_{lm}\mu_k^{lm},
\end{align}
we have
\begin{align}
\label{eq:mergeStateRemove}
\mathsf{\hat{\boldsymbol{x}}}_{k|k} = & \mathsf{\hat{\boldsymbol{x}}}_{k|k}^{ij},
\\ \label{eq:mergedCovRemove} \mathsf{P}_{k|k}  = & \mathsf{P}_{k|k}^{ij} .
\end{align}

As mentioned before, removing the clusters with smaller weights is less computationally complex than merging the clusters to one for two reasons. First, by determining the active model before evaluating the parameters of all clusters (if possible), computational resources can be saved. Second, evaluating the moments of the GM posterior, i.e.~\eqref{eq:mergedState}--\eqref{eq:mergedCov} is more computationally complex than simply taking the moments of the active model as in \eqref{eq:mergeStateRemove}--\eqref{eq:mergedCovRemove}.

Besides the computational complexity, finding and using the active model can provide a better approximation than merging the clusters. This is due to the fact that by making a soft decision about the active model we are drifting from the true distribution. Specifically, since at each iteration there is only one active model, by including the other clusters in the estimation we are adding bias to the estimated state. Ideally, if the active model can be determined correctly, the lower bound on MSE can be reached~\cite{flam_mmse_2012}. However, if the active model cannot be determined correctly, merging the clusters to one can yield better estimation accuracy and precision. In this work we propose a reduction scheme which determines the active model and removes the other clusters. We show through simulation that our proposed method for determining the active cluster, can show a better performance than the method described in~\eqref{eq:determineActive}--\eqref{eq:mergedCovRemove}.

In the following we use GSF-merge to refer to the filter using the first reduction scheme, i.e.~\eqref{eq:mergedState}--\eqref{eq:mergedCov} and we denote the filter using the second reduction scheme, i.e.~\eqref{eq:mergeStateRemove}--\eqref{eq:mergedCovRemove} by GSF-remove.

For comparison purposes we also define \textit{Matched filter} as the filter which removes all but the active cluster. Since this requires having prior information about the active model, it cannot be used in practice and it is only implemented in simulations for comparison purposes. Matched filter is proved to have the lower bound on MSE of the GSF~\cite{flam_mmse_2012}.


\section{The Proposed Reduction Algorithm}
\label{sec:myCont}

\begin{figure}[!t]
\centering
\includegraphics[width=3.0in]{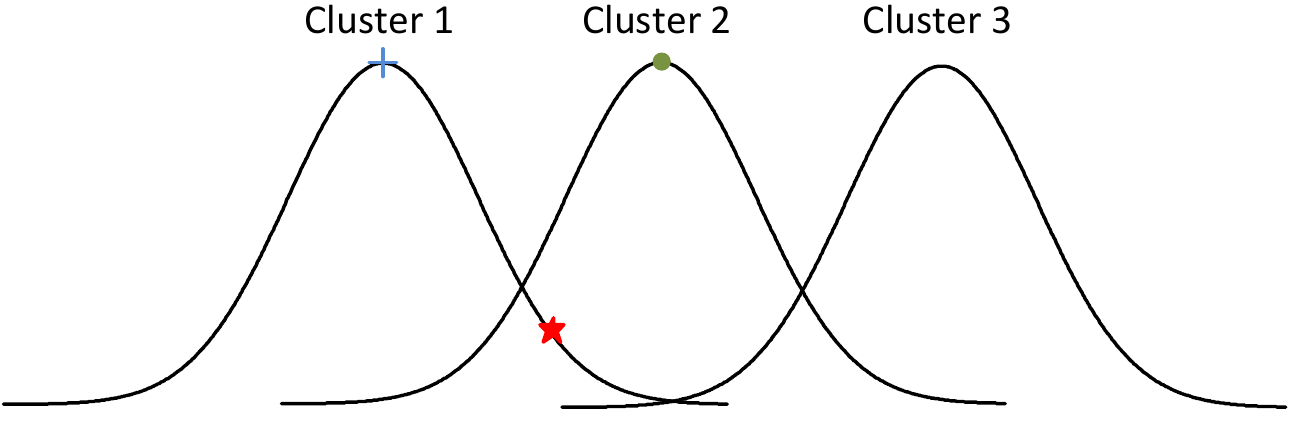}
\caption{A schematic depiction of the case where the Matched filter will not provide the most accurate and precise estimation. The red star shows the true state which belongs to Cluster 1. However, the mean of Cluster 1, indicated by the blue \(+\) provides a less accurate estimation when compared with the non-matched Cluster 2.}
\label{fig:SchematicFig}
\end{figure}

In this section, we propose a method for determining the cluster which provides the most accurate estimation. Our proposed method relies on the fact that if the state vector is known, the noise vectors can be evaluated and the posterior cluster closest to the corresponding noise vectors can be determined. It is worth noting that this cluster can be different from the cluster corresponding to Matched filter. This is depicted schematically in Fig.~\ref{fig:SchematicFig}. Hence, in theory, this method can provide an estimator which has an MSE lower than the MSE of Matched filter, i.e. the lower bound of MSE~\cite{flam_mmse_2012}. However, since the true state vector is unknown we rely on an initial state estimation, denoted by \(\check{\mathbf{x}}_k\), to approximate the noise vectors and determine this cluster.

Using the initial state estimation, \(\check{\mathbf{x}}_k\), in~\eqref{eq:process}--\eqref{eq:measurement}, we can find an approximation for the process noise, \(\check{\mathbf{v}}_k\), and the measurement noise, \(\check{\mathbf{w}}_k\), as follows:
\begin{align}
\label{eq:findvk}
\check{\mathbf{v}}_k  =& \check{\mathbf{x}}_k - F_{k}\hat{\mathbf{x}}_{k-1},\\
\label{eq:findwk}
\check{\mathbf{w}}_k = & \mathbf{z}_k - H_k\check{\mathbf{x}}_{k}.
\end{align}
Now, the noise approximations can be used to determine the noise clusters in~\eqref{eq:priorDist}--\eqref{eq:likelihoodDist} which are more likely, i.e.
\begin{align}
\nonumber
M^* =&\arg \max_{ij} w^i_k \mathcal{N}\left(\hat{\mathbf{v}}_k; \mathbf{u}^i_k, Q^i_k \right) \\
\label{eq:FindM}
&\times p^j_k \mathcal{N}\left(\hat{\mathbf{w}}_k; {\mathbf{b}}^j_k, R^j_k \right).
\end{align}

Based on the above, it is evident that the performance of the proposed reduction scheme relies on the accuracy of the noise approximations, which is dependent on the initial state estimation. Hence, it is important to choose an initial estimation that is accurate enough to be used in~\eqref{eq:findvk}--\eqref{eq:findwk}, and at the same time it is less computationally complex than the other methods. In the following, we propose five methods for evaluating the initial state estimation. The main difference between these methods is their computational complexity. In Section~\ref{sec:SimResults} we compare the performance of using these initial state estimations in our proposed reduction scheme for different posteriors, through computer simulations.

\subsection{Choosing the Initial State Estimation}
\label{sec:InitialStateEstimation}
\subsubsection{GSF-merge state estimation (Red-GSFM)}
\label{sec:UsingGSF-merge}
In this case, the computational complexity of the reduction scheme is very close to that of GSF-merge, as the parameters of all filters are evaluated. However, since only the state estimation is used, there is no need to evaluate~\eqref{eq:mergedCov}, and this makes the computational complexity of the proposed scheme slightly better.

\subsubsection{GSF-remove state estimation (Red-GSFR)}
\label{sec:UsingGSF-remove}
In this case, the computational complexity of the proposed reduction scheme is similar to GSF-remove.

\subsubsection{Using preloaded Kalman gains (Red-PKG)}
\label{sec:UsingPreloadedKal}
In this case, instead of evaluating the gains of individual filters in GSF, we use preloaded Kalman gains. Since the Kalman filter's gains can be evaluated offline, \(\check{\mathbf{x}}_k\) can be determined at a lower computational complexity. In other words, we use GSF-merge state estimation, but with preloaded Kalman gains. Hence, there is no need to evaluate the gains of individual filters to find \(\check{\mathbf{x}}_k\). 

\subsubsection{Using steady state gains (Red-SSG)}
\label{sec:UsingSteadyStateGains}
This is similar to Red-PKG, but using the steady state gains of individual filters instead of the preloaded Kalman gains. Since the steady state gains of the individual filters in GSF can be precomputed and stored, using this method has a lower computational complexity when compared to using the state estimation of GSF-merge.

\subsubsection{Deriving and using Kalman gain (Red-DKG)}
In this case, instead of preloading the gains (either steady state or Kalman gains), we evaluate and use the gain of a Kalman filter run at that iteration. This is different from Red-PKG, as the previous iteration parameters are different due to using the proposed reduction scheme. This method does not require precomputing and storing gains, but requires the evaluation of Kalman parameters instead. 

\section{Simulation Results}
\label{sec:SimResults}
In this section we apply our proposed reduction scheme with the initial state estimations described in Section~\ref{sec:InitialStateEstimation}, and compare these filters with Kalman filter, GSF-merge, and GSF-remove. We use two simulation scenarios: A. With synthetically generated data; and B. With experimental data gathered from indoor localization system with Ultra-WideBand (UWB) sensors. For the sake of consistency, in both scenarios we assume a 2D localization problem, where the state vector contains the location information of a mobile indoor target. Hence, we use the same process and measurement equations for both cases with  
\begin{align}
F_k = & \begin{bmatrix}
1 & \Delta t_{k} \\
0 & 1
\end{bmatrix},\\
H_k = & \begin{bmatrix}
1 & 0
\end{bmatrix},
\end{align}
where \(\Delta t_{k}\) is the time interval between the measurements \(z_{k-1}\) and \(z_k\) and is a multiple of \(0.1080\)~s.\footnote{In the indoor location system used, the measurements are received at these intervals.} The motion model we use is a random walk velocity motion model, hence,
\begin{align}
\mathbf{v}_{k} = v_k \times \begin{bmatrix}
\Delta t_{k} \\
1
\end{bmatrix},
\end{align}
where \(v_k\) is a univariate GM random variable with \(C_{v_k}\) clusters, with \(\left\lbrace {u}^i_k,1 \leq i \leq C_{v_k}\right\rbrace\) the cluster means, \(\left\lbrace {\sigma^i}^2_k,1 \leq i \leq C_{v_k}\right\rbrace\) the cluster variances. Hence, the parameters in~\eqref{eq:priorDist} are evaluated as:
\begin{align}
\forall i; \; 1 \leq i \leq C_{v_k}, \; & \mathbf{u}^i_k = {u}^i_k \times \begin{bmatrix}
\Delta t_{k} \\
1
\end{bmatrix},\\
& Q^i_k = {\sigma^i}^2_k \times \begin{bmatrix}
\Delta t_{k}^2 &  \Delta t_{k}\\
\Delta t_{k} & 1
\end{bmatrix}.
\end{align}
%
%

We use Root-Mean-Square Error (RMSE) and Circular Error Probable (CEP), to evaluate the accuracy and precision of the estimators, respectively. If we define \(\epsilon_k\) as the estimation error at iteration \(k\), we have:
\begin{align}
\textrm{RMSE} \triangleq & \left({\frac{1}{N}\sum \limits_{k=1}^{N}{\epsilon_k}^2}\right)^{1/2}, \\
\textrm{CEP} \triangleq & F^{-1}_{|\epsilon|} \left(0.5\right),
\end{align}
where \(N\) is the total number of iterations and \(F^{-1}_{|\epsilon|}\) represents the inverse cdf of error evaluated over the whole experiment time-span.

\subsection{Synthetic Models}
In our synthetic simulation scenario we use the same GM distribution for both process and measurement noises. 
We use three different models for this purpose:
\begin{itemize}
{\setlength\itemindent{35pt}\item[\textbf{Model 1:}] Symmetric distribution with the mixands all having the same coefficients.}
{\setlength\itemindent{35pt}\item[\textbf{Model 2:}] Symmetric distribution with mixands possibly having different weights.}
{\setlength\itemindent{35pt}\item[\textbf{Model 3:}] Asymmetric distribution.}
\end{itemize}
We use GM distributions with 5 clusters, each having a variance of \(1\), i.e. we have
\begin{align}
 C_{v_k}  = C_{w_k} & = 5, \\
\forall i; \;  1 \leq i \leq 5, & \; {\sigma^i}^2_k = 1 ,
\\ \forall j; \;  1 \leq j \leq 5, & \;  R^j_k = 1.
\end{align}
The coefficients, \(\mathbf{w}\), and the means, \(\mathbf{m}\), of the clusters are given in Table~\ref{tab:GMModelsparams}, where
\begin{align}
\mathbf{w}  & = \begin{bmatrix}
w_k^1, \cdots, w_k^5
\end{bmatrix} = \begin{bmatrix}
p_k^1,\cdots , p_k^5
\end{bmatrix}, \\
\mathbf{m} & = \begin{bmatrix}
b_k^1,\cdots , b_k^5
\end{bmatrix} = \begin{bmatrix}
u_k^1,\cdots , u_k^5
\end{bmatrix}. 
\end{align}
To compare the effect of multimodality on the different filtering schemes, the parameter \(c\) is used to change the distance between the means of the clusters. This is measured by approximating the Kullback-Leibler (KL) divergence\footnote{Using Monte-Carlo simulations with \(2\times 10^5\) samples.} between the GM noise distribution and its corresponding moment-matched Gaussian. 

\begin{table}[!t]
\centering
\caption{The parameters of the GM models used for generating synthetic data}
\label{tab:GMModelsparams}
\begin{tabular}{c l}
\hline
\textbf{Model 1} & \(\mathbf{w} = \left[0.2,0.2,0.2,0.2,0.2\right]\) \\
& \(\mathbf{m}=c \left[-50,-30,0,30,50\right]\) \\
\hline
\textbf{Model 2} & \(\mathbf{w} = \left[0.1,0.1,0.6,0.1,0.1\right]\) \\
& \(\mathbf{m}=c\left[-50,-30,0,30,50\right]\) \\
\hline
\textbf{Model 3} & \(\mathbf{w} = \left[0.5,0.1,0.1,0.1,0.2\right]\) \\
& \(\mathbf{m}=c\left[-50,10,30,50,80\right]\) \\
\hline
\end{tabular}
\end{table}

\begin{figure}[!t]
\centering
\includegraphics[width=3.5in]{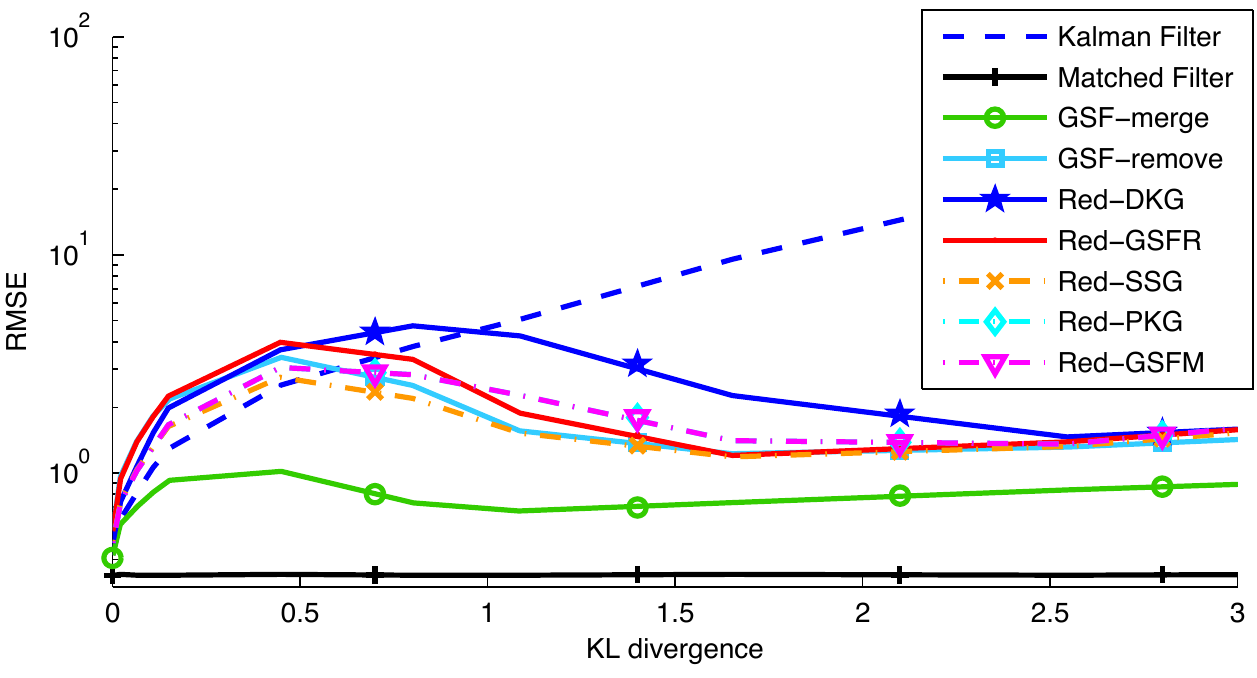}
\caption{RMSE for the synthetic data generated using Model 1 vs. KL divergence between the noise distribution and the moment-matched Gaussian pdf.}
\label{fig:RMSEM1}
\end{figure}

\begin{figure}[!t]
\centering
\includegraphics[width=3.5in]{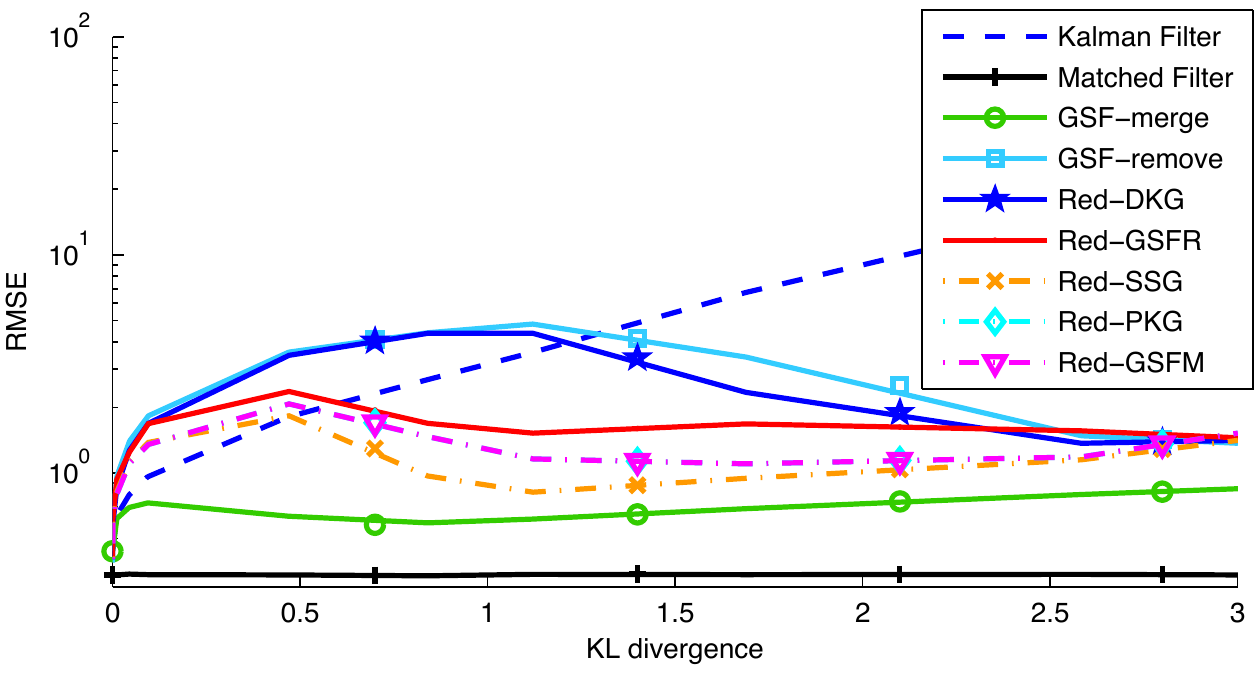}
\caption{RMSE for the synthetic data generated using Model 2 vs. KL divergence between the noise distribution and the moment-matched Gaussian pdf.}
\label{fig:RMSEM2}
\end{figure}

\begin{figure}[!t]
\centering
\includegraphics[width=3.5in]{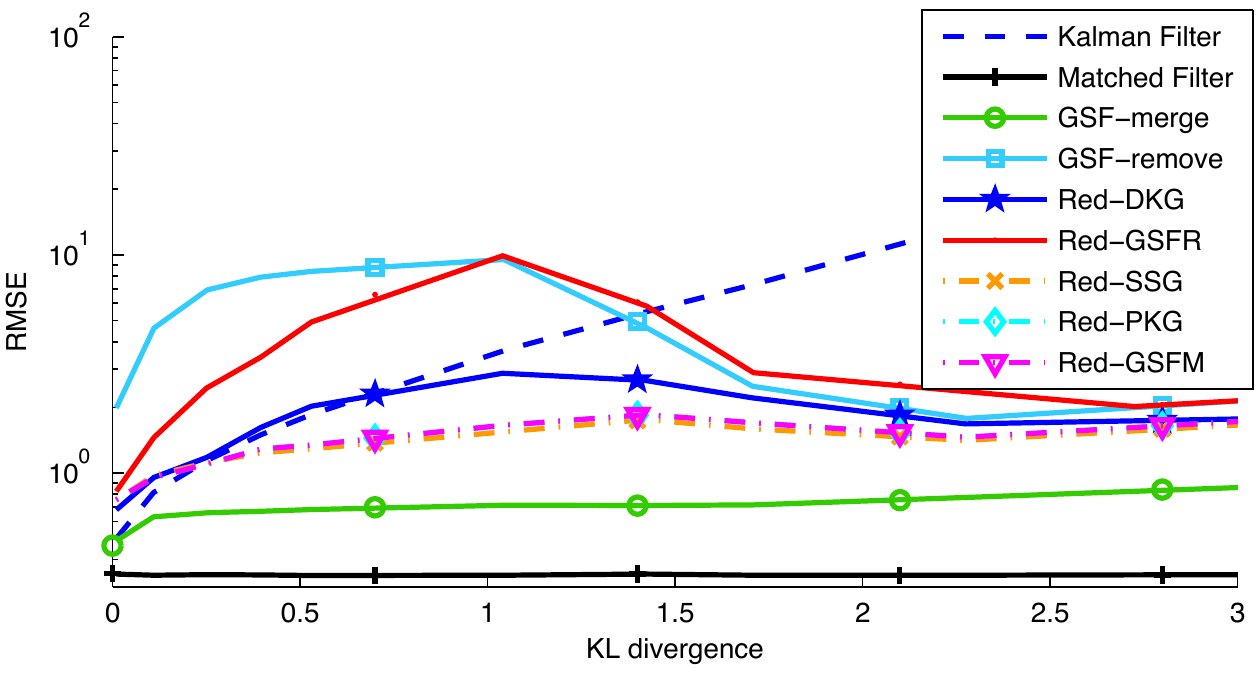}
\caption{RMSE for the synthetic data generated using Model 3 vs. KL divergence between the noise distribution and the moment-matched Gaussian pdf.}
\label{fig:RMSEM3}
\end{figure}

The RMSE of the different filtering schemes are given in Figures~\ref{fig:RMSEM1}--\ref{fig:RMSEM3} for different KL divergences.\footnote{\textcolor{CorCol}{Since the behavior of CEP is similar to RMSE for synthetic data, we did not include the CEP results.}} The values depicted in these figures, are the average of \(1000\) Monte-Carlo runs to achieve \(\%95\) confidence interval. Matched filter results are depicted for comparison purposes only since it provides the lower bound on MSE for GSF. 

The KL divergence between the noise distribution and its corresponding moment-matched Gaussian distribution can be used as a measure of how well the noise distribution can be approximated by a single Gaussian. Hence, for small values of KL divergence (\(\textrm{KL} \leq 1\) for Models 1--2 and \(\textrm{KL} \leq 1.5\) for Model 3), Kalman filter provides very good estimations both in terms of accuracy and precision. However, as we increase the KL divergence, the performance of Kalman filter as a state estimator drops, while the other methods show improved estimation accuracy and precision. Since the variance of all clusters is the same in all models, it is easy to see that the preloaded Kalman gains and the gains of individual filters will be the same for all models
. Consequently, Red-GSFM and Red-PKG will provide the same results. 

\begin{figure}[!t]
\centering
\includegraphics[width=3.5in]{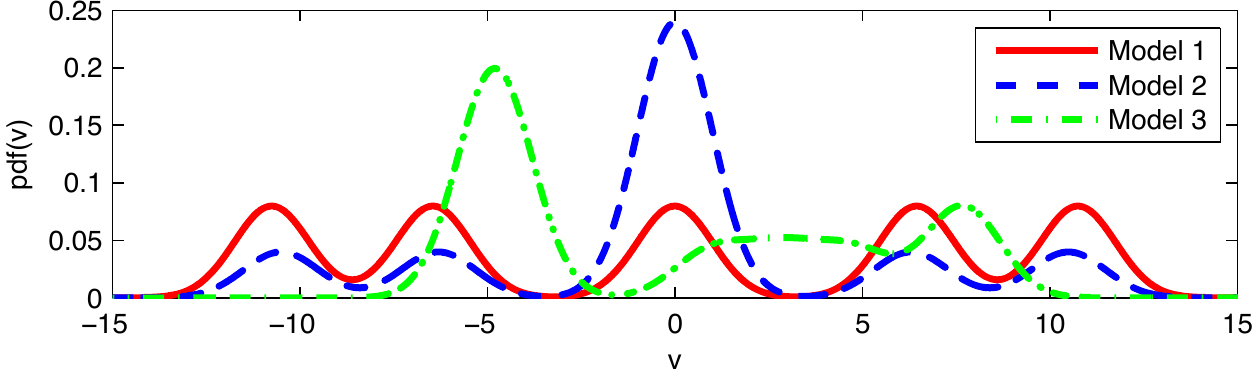}
\caption{pdf of noise for the three synthetic models is depicted. We use \(c = 0.21\), \(c = 0.215\) and \(c = 0.096\) for Model 1,2, and 3, respectively, to have an approximate KL divergence of \(0.5\).}
\label{fig:noiseDists}
\end{figure}

As shown earlier, the performance of our proposed scheme with different initial state estimations depends on the accuracy of the initial estimations. Hence, depending on the shape of the noise models, the methods in Section~\ref{sec:InitialStateEstimation} have different accuracies. To show this more clearly, in Fig.~\ref{fig:noiseDists} we provide the pdf of noise for different models at \(\textrm{KL} \approxeq 0.5\). Based on this figure, the noise distribution in Model 1 shows clear multi-modality, hence a single Gaussian distribution cannot represent it accurately enough. This is why for Model 1 Red-DKG has a lower performance when compared with other reduction schemes. For Model 2, since the noise distribution is a symmetric pdf with a more significant mode at the center, the performance of Red-DKG and Red-GSFR are very similar. Specifically, due to the larger weight of the cluster in the center, the other clusters are more likely to be removed. Additionally, the moment-matched Gaussian distribution of this noise model has a mode at the center. Thus, the two methods have similar state estimations. On the other hand, due to the shape of the noise distribution (having a significant mode), Red-GSFR has a good performance since GSF-remove can provide a good initial state estimation. Contrarily, Model~3 has an asymmetric distribution with two more significant modes. Hence, GSF-remove cannot provide a good state estimation in this model, and consequently Red-GSFR does not perform well, either. Based on the results it is evident that depending on the shape of the noise distributions, suitable initial state estimation can be chosen and the proposed reduction scheme can improve the performance of estimation. 

%
%

\subsection{Experimental model}
\begin{figure}[!t]
\centering
\includegraphics[width=2.8in]{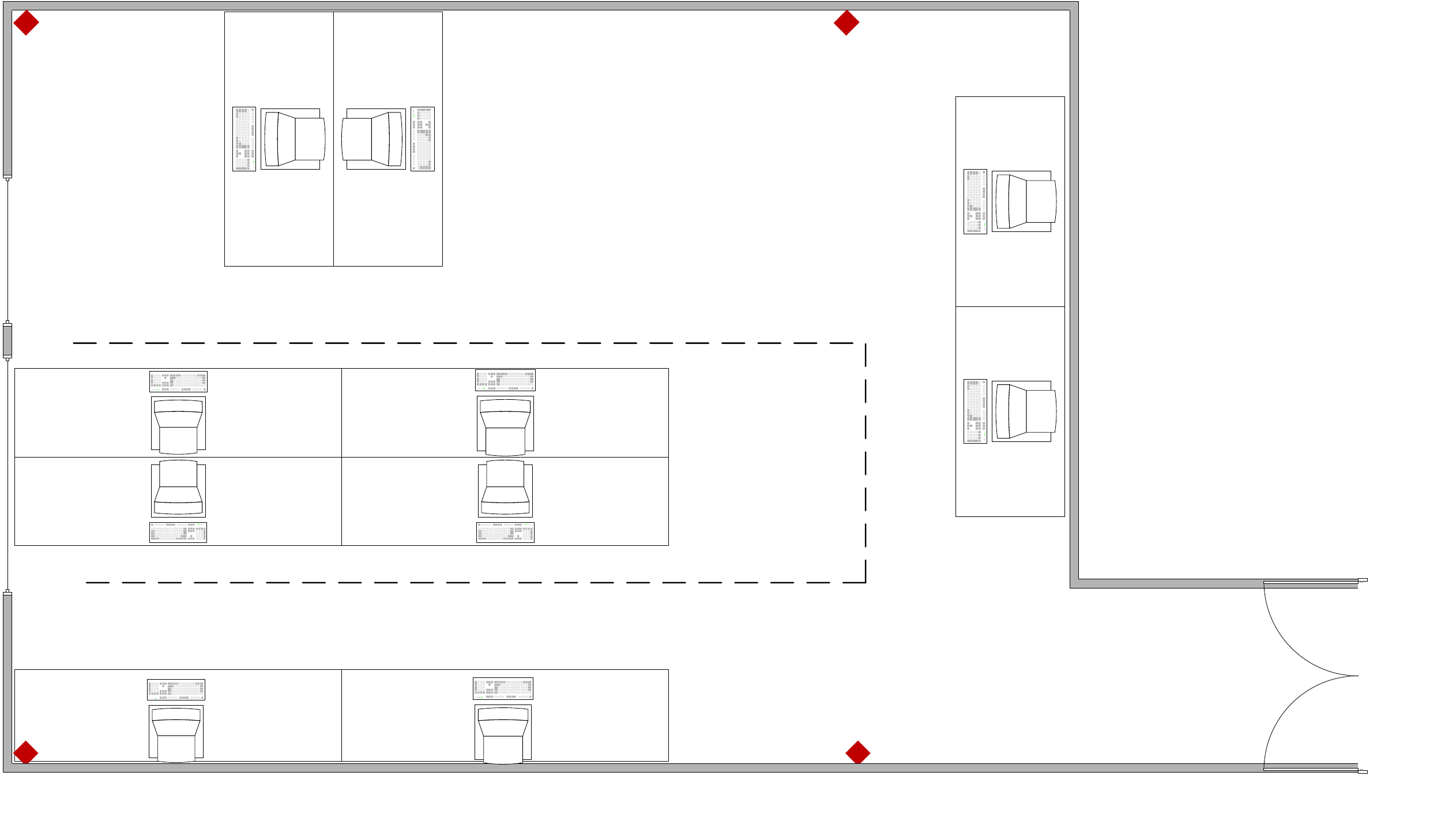}
\caption{A floor plan of the experimental simulation setup is shown, with the dashed line indicating the trajectory of the moving target, and the red diamonds showing the location of UWB receivers.}
\label{fig:SimMap}
\end{figure}

\begin{figure}[!t]
\centering
\subfloat{\includegraphics[width=3.2in]{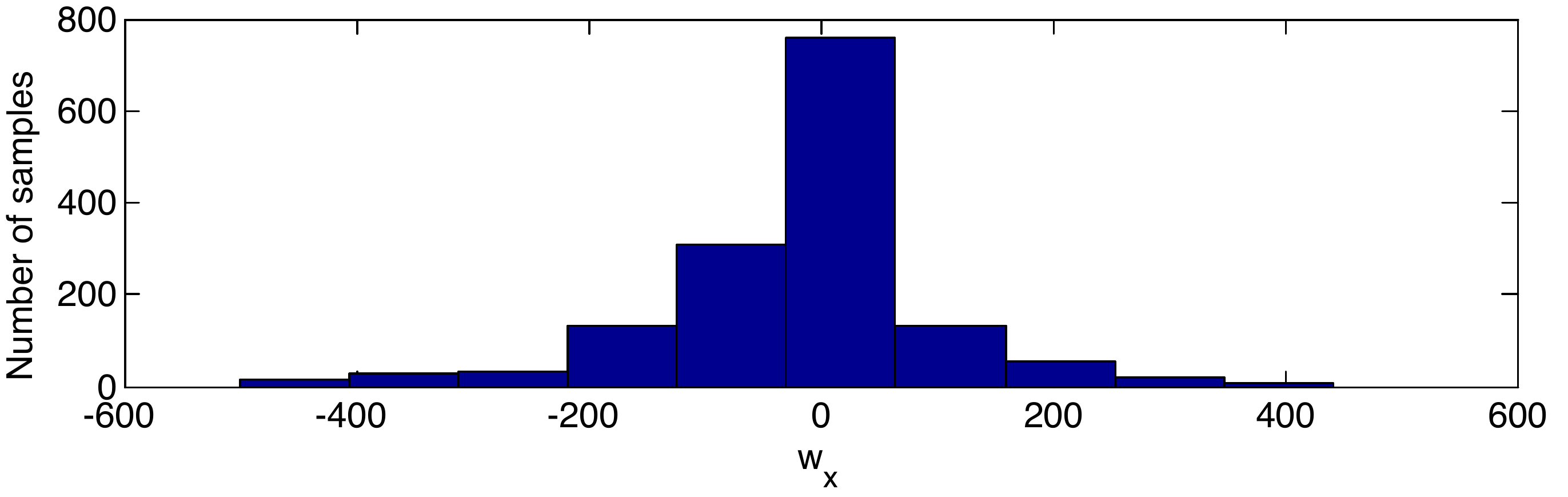}}%
\hfill
\subfloat{\includegraphics[width=3.2in]{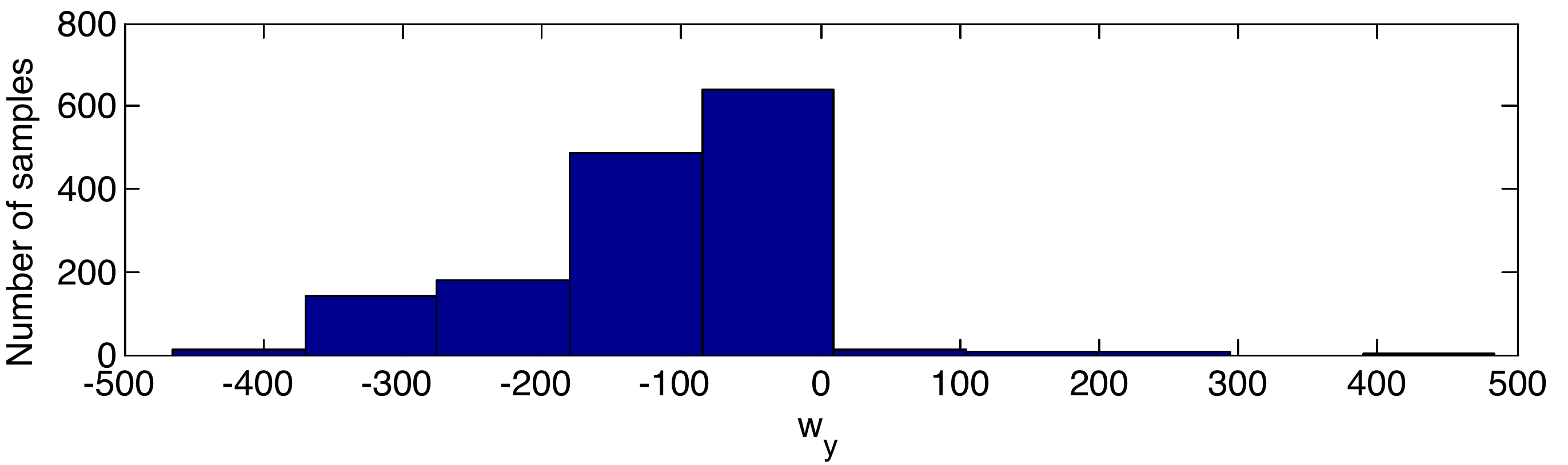}}
\caption{Histogram of measurement noise in \(x\) and \(y\) directions}
\label{fig:ExperMeasurementNoisepdf}
\end{figure}

Our experimental indoor localization system is composed of \textit{Ubisense}~\cite{website:ubisense} UWB location sensors and receivers. We have four UWB receivers, four stationary object and a moving target (Fig.~\ref{fig:SimMap}). The location is estimated by computing the time and angle difference of arrival of UWB pulses that are sent by a UWB tag attached to the objects. However, due to large estimation errors, the location information should be post-processed. Using the location information of the stationary objects, the histograms of measurement noise in \(x\) and \(y\) directions are depicted in Fig.~\ref{fig:ExperMeasurementNoisepdf}.

\begin{figure}[!t]
\centering
\subfloat{\includegraphics[width=3.2in]{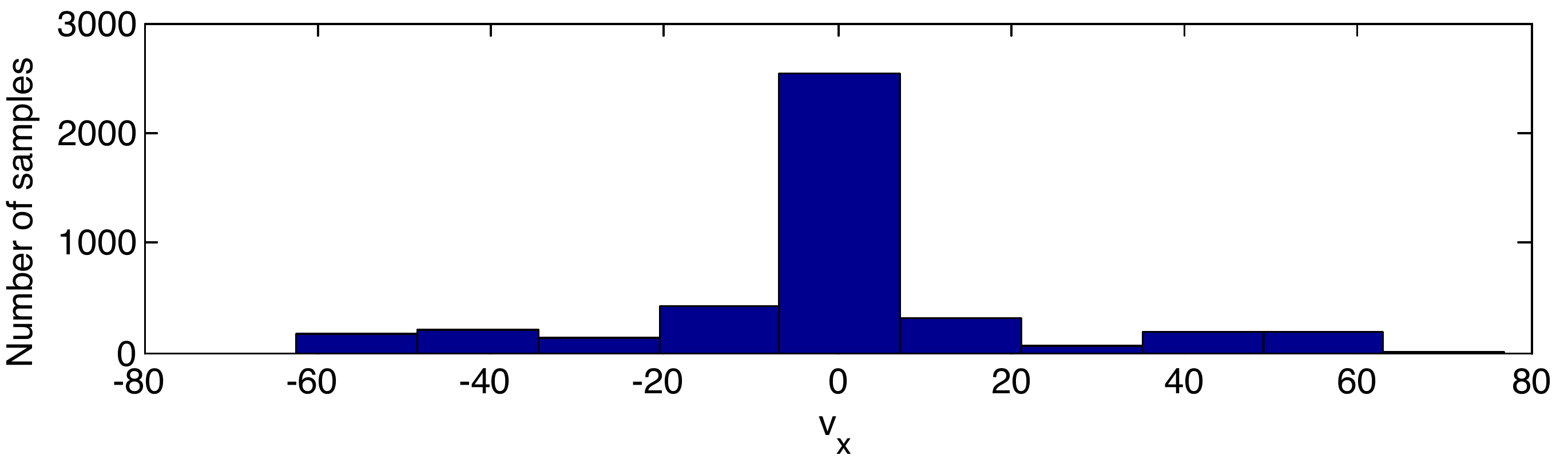}}%
\hfill
\subfloat{\includegraphics[width=3.2in]{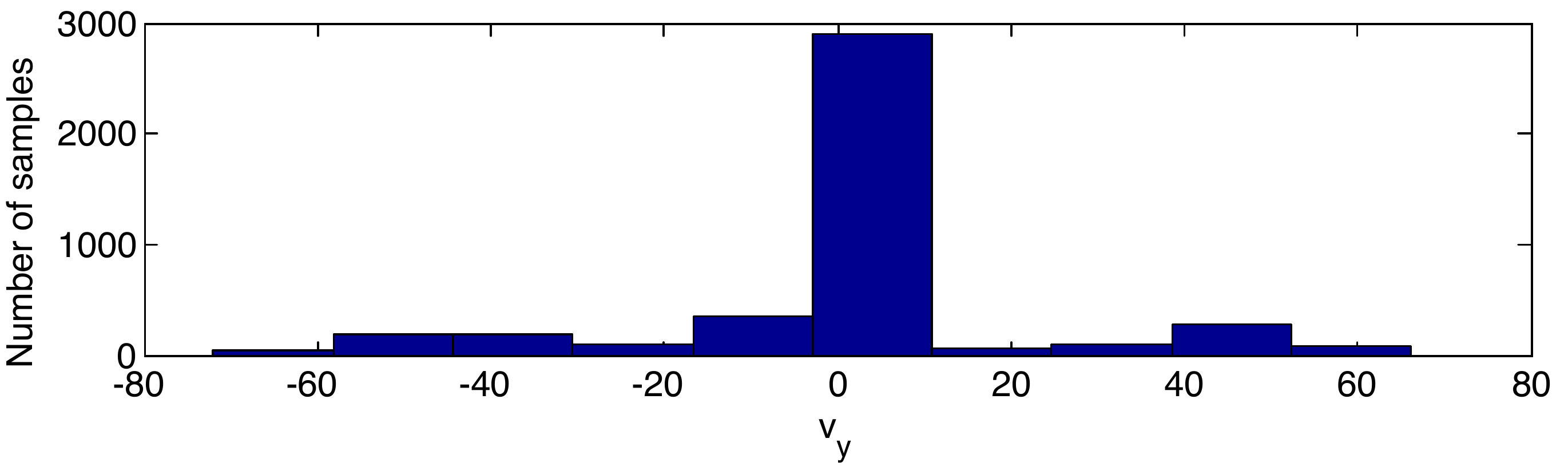}}
\caption{Histogram of process noise in \(x\) and \(y\) directions}
\label{fig:ExperProcessNoisepdf}
\end{figure}

Since the trajectory and the ground truth about the location of the mobile object is known, it can be used to approximate the process noise pdfs. Fig.~\ref{fig:ExperProcessNoisepdf} shows the histograms of process noise in \(x\) and \(y\) directions for 20 experiments. 

Using the noise histograms, they can be approximated with GM distributions. Table~\ref{tab:KLdiv} shows the parameters of the GM noise distributions and the KL divergence with their corresponding moment-matched Gaussian pdf, in \(x\) and \(y\) directions. In this table, we use \(\prescript{}{}{[a^i]}^{n}_{1}\) to denote \([a^1,\cdots,a^n]\).

\begin{table}[!t]
\centering
\caption{\textcolor{CorCol}{The parameters of the GM noise distributions}}
\label{tab:KLdiv}
\begin{tabular}{@{\hspace{0.1cm}}l@{\hspace{0.1cm}}|c@{\hspace{0.1cm}}|@{\hspace{0.1cm}}c@{\hspace{0.1cm}}|l}
                        &  \multirow{2}{*}{Noise}    & KL  & \multirow{2}{*}{Parameters} \\
                        & & divergence & \\
                       \hline
\normalsize 
{\multirow{12}{*}{\(x\)}} & \multirow{6}{*}{\(v_k\)} & \multirow{6}{*}{\(0.4253\)} & {\(C_{v_k}=3\)}
\\[1mm]
& & & {\(\prescript{}{}{[w_k^i]}^{3}_{1} = [0.13,0.77,0.099]\)}       \\[1mm]
& & &  {\(\prescript{}{}{[u_k^i]}^{3}_{1} = [-41.44,	0.51,	49.79]\)} \\[1mm]
& & &  {\(\prescript{}{}{[{\sigma^i}^2_k]}^{3}_{1} = [148.24,48.38,	83.75]\)} \\[1mm]
\cline{2-4}
& \multirow{6}{*}{\(w_k\)} & \multirow{6}{*}{\(0.1759\)} & {\(C_{w_k}=3\)} \\[1mm]
& & &  {\(\prescript{}{}{[p_k^j]}^{3}_{1} = [0.07,	0.85,	0.08]\)}       \\[1mm]
& & & {\(\prescript{}{}{[b_k^j]}^{3}_{1} = [-300.01,	-17.06,	207.37]\)} \\[1mm]
& & & {\(\prescript{}{}{[R_k^j]}^{3}_{1} = [8163.20,	3611.99,	5677.21]\)} \\[1mm]
\hline                       
\multirow{16}{*}{\(y\)} & \multirow{8}{*}{\(v_k\)} & \multirow{8}{*}{\(1.1971\)}       & \(C_{v_k}=9\) \\[1mm]
& & &  \(\prescript{}{}{[w_k^i]}^{9}_{1} = [ 0.01,	0.06,	0.03, 0.03,\)       \\
& & & \( \quad \quad \quad \quad \; 	0.72,	0.04 ,  0.02,	0.06,	0.03 ]\)\\
& & &  \(\prescript{}{}{[u_k^i]}^{9}_{1} = [-63.38,	-48.73,	-35.65,	-17.40,\) \\ 
& & &\(\quad \quad \quad \quad \; 	-0.32,	9.52,	30.09,	44.24,	54.35]\) \\
& & &  \(\prescript{}{}{[{\sigma^i}^2_k]}^{9}_{1} = [24.34,	21.53,	18.18,	23.62,\) \\ 
& & &\(\quad \quad \quad \quad \; 	3.13,	12.16,	18.81,	12.96,	15.44]\) \\
\cline{2-4}
& \multirow{6}{*}{\(w_k\)} & \multirow{6}{*}{\(0.0200\)} & \(C_{w_k}=2\) \\[1mm]
& & &  \(\prescript{}{}{[p_k^j]}^{2}_{1} = [0.98,	0.02]\)\\[1mm]
& & & \(\prescript{}{}{[b_k^j]}^{2}_{1} = [-125.93,	147.25]\) \\[1mm]
& & & \(\prescript{}{}{[R_k^j]}^{2}_{1} = [8500.19,10809.10]\) \\[1mm]
\end{tabular}
\end{table}

The results for the experimental model (averaged over 20 experiments) are provided in Table~\ref{tab:Experexperdata}. Since the KL divergences between the noise distributions and their corresponding moment-matched Gaussian density is small, Kalman filter provides good estimations both in terms of accuracy and precision. However, our proposed method (with a suitable scheme to find the initial state estimation) has the best performance in both directions. 


\begin{table}[!t]
\centering
\caption{RMSE and CEP for synthetic data gathered from the indoor localization system}
\label{tab:Experexperdata}
\begin{tabular}{l|l|l|l|l|}
     \multicolumn{1}{l}{ }   & \multicolumn{2}{c}{x}        & \multicolumn{2}{c}{y} \\ \cline{2-5}
           & \multicolumn{1}{l|}{RMSE}      & CEP     & RMSE      & CEP       \\ \hline
\multicolumn{1}{|l|}{Kalman}                                      & 88.1850    & 66.3306   & 73.0512    & 60.4984    \\ \hline
\multicolumn{1}{|l|}{GSF-merge}                                   & 85.7485    & 55.2261   & 74.3982    & 61.7358    \\ \hline
\multicolumn{1}{|l|}{GSF-remove}                                  & 110.4006   & 91.7458   & 75.1093    & 48.6651    \\ \hline
\multicolumn{1}{|l|}{Red-GSFR} & 94.4281    & 59.8799   & \textbf{71.4266}    & \textbf{35.1540}    \\ \hline
\multicolumn{1}{|l|}{Red-GSFM}  & 84.7763    & \textbf{49.9824}   & 78.0184    & 44.7451    \\ \hline
\multicolumn{1}{|l|}{Red-PKG}      & 84.6138    & 53.4104   & 212.2485   & 145.8093   \\ \hline
\multicolumn{1}{|l|}{Red-SSG}          & \textbf{75.9668}    & 51.5365   & 92.3020    & 73.6679    \\ \hline
\multicolumn{1}{|l|}{Red-DKG}                & 81.9455    & 50.8301   & 86.7781    & 56.4840    \\ \hline
\end{tabular}
\end{table}

\section{Conclusion} 
\label{sec:Conclusion}
In this work we propose a low computational complexity reduction scheme for Gaussian Sum Filters (GSF) in linear dynamic state space systems with Gaussian Mixture (GM) noise distributions. Our method relies on the fact that at each iteration, only one of the clusters of the GM noise distributions are \textit{active}, and uses simple a posteriori probability comparisons to find this active model. This is done by using an initial state estimation to approximate the noise vectors. Hence, the performance of the proposed reduction scheme is dependent on the accuracy of the initial state estimation. We propose five different methods to find the initial state estimation and compare their performances for different noise distributions through simulations. The simulation results show that our proposed reduction scheme can perform better with suitable choice of the initial state estimation. 

\section*{Acknowledgment}
This work was partly supported by the Natural Sciences and Engineering Research Council (NSERC) and industrial and government partners, through the Healthcare Support through Information Technology Enhancements (hSITE) Strategic Research Network.




%

\bibliographystyle{IEEEtran}
\bibliography{KFBank}

%

%






\end{document}